\documentclass[preprint,12pt,number]{elsarticle}

\usepackage{graphicx}
\usepackage{epsfig}
\usepackage{amssymb}

 \journal{PhysicaD:Nonlinear Phenomena}

\begin{document}

\begin{frontmatter}



\title{\textbf{Chaos at Nonlinear NMR}}


\author[a]{A.Ugulava}
\author[a]{S.Chkhaidze\corref{*}}
\cortext[*]{Corresponding author}
 \ead{s.chkhaidze@tsu.ge}
\author[a,b]{L.Chotorlishvili}
\author[c]{Z.Rostomashvili}
\address[a]{I.Javakhishvili Tbilisi State University,
I.Chavchavadze av., 3, Tbilisi 0128, Georgia}
\address[b]{Institute for Physik, UniversitÄat Augsburg, 86135 Augsburg, Germany}
\address[c]{I. Gogebashvili Telavi State University,
University street, 1, Telavi 2200, Georgia}

\begin{abstract}
The hodographs of magnetization of nonlinear NMR are investigated
in the conditions of resonance on the unshifted frequency. It is
shown that, depending on the value of amplitude of the variable
field and value of frequency shift, topologically different
hodographs separated from each other by separatrix are obtained.
It is shown that the set of hodograph points, being obtained by
the stroboscopic method, is chaotic and the change of
$z$-component of magnetization has the form of solitons
chaotically changing the sign.
\end{abstract}


\end{frontmatter}


\section{Introduction}
\label{1}

Investigations of nonlinear magnetic resonance in magnetically
ordered materials have been carried out for a long time [1],
however the interest to them is still increasing due to discovery
of new physical phenomena in magnetics [2--9]. On the other hand
the interest to chaotic Hamiltonian systems does not weaken
[10--12]. That is why the study of any questions on the butt of
these two fields (our work concerns to) in our opinion is quite
topical. In particular most of spin systems considered as
promising candidates for applications in quantum computation are
nonlinear [13--17]. Nonlinearity is a principle problem for
controlled states. Note that for a linear system in a constant
magnetic field, magnetization vector can be deflected by applying
a variable radio frequency field. However for nonlinear systems,
situation is completly different.

In magnetically ordered crystals at quite, low temperatures a
shift of frequency precession of nuclear magnetization ${\bf m}$
proportional to the longitudinal component $m_z$ (dynamical shift)
origins. In the NMR conditions appearance of dynamic shift leads
to essentially nonlinear character of the motion ${\bf m}$. For
example, in [18] it was shown, that fulfilling the conditions
$\omega_1=\omega_p$, where $\omega_p$ is equilibrium value of
dynamic shift, $\omega_1=\eta\gamma H_1,~H_1$ is an amplitude of
linearly polarized variable field, applied in the transverse
plane, $\eta$ is gain factor, $\gamma$ is a gyromagnetic ratio,
the motion of ${\bf m}$ vector in the rotational coordinate system
becomes aperiodic (period $T\rightarrow\infty$). Trajectory, along
which the end of ${\bf m}$ vector moves (hodograph) in this case
is a separatrix topologically separating differing trajectories.
It will be shown below that points put according to hodograph
being close to separatrix form a stochastic set. Simple, but
effective method of finding the stochastic motion (stroboscopic
method) is offered.

Peculiarity of dynamical systems is extreme sensitivity of motion
near separatrix, even with respect to the slightly change of
initial conditions or adiabatic perturbation. Purpose of our paper
is demonstration at the fact that control of the state of
nonlinear system is not possible near separatrix. In order to
proof this statement, we shall consider realistic physical system.
Namely, we shall study motion of nuclear magnetization in the
presence of large dynamical shift of frequency.

\section{\textbf{Equations of motion of magnetization at nonlinear NMR}}
\label{2}

Let us direct the axis z along the total field ${\bf H}={\bf
H}_0+{\bf H}_{loc}$, where ${\bf H}_0$ and ${\bf H}_{loc}$ are
respectively external and internal magnetostatic fields on the
nuclei. Let us assume that the variable fields also effect on the
nuclear spin-system

\begin{equation}
{\bf H}_1(t)={\bf H}_1\cos\omega t,~~~{\bf H}_1\perp{\bf H}
\end{equation}
and
\begin{equation}
{\bf h}_1(t)={\bf h}_1\cos\nu t,~~~{\bf h}\parallel{\bf H}
\end{equation}
Directing ${\bf H}_1$ along the axis $x$, the system of nonlinear
differential equations describing the motion of nuclear
magnetization with the account of dynamical shift of frequency
precession in the rotating system of coordinates around the axis
$z$ with the frequency can be written as
follows\footnote{Investigation of Bose-Hubbard model is also
reduced to consideration of such type of equations [19]}
$$\dot{x}=(\Delta -\omega_pz)y+\nu_1y\cos\nu t,$$

\begin{equation}
\dot{y}=-(\Delta -\omega_pz)x+\frac{1}{2}\omega_1z-\nu_1x\cos\nu
t,
\end{equation}

$$\dot{z}=-\frac{1}{2}\omega_1y,~~\nu_1=\gamma h,~~\Delta =\omega-\omega_0,~~ \omega_0=\gamma H.$$
Here we introduced dimensionless components of magnetization
$\alpha=m_{\alpha}/m,$  where $m=|{\bf m}|,~\alpha=x,y,z.$ Our
observation neglects relaxation effects and magnetostatic
inhomogeneity of the local field is valid in the case of fields
with duration $\tau\ll T_1,~T_2,~2\pi/\Omega,$ where $T_1$ and
$T_2$ are the times of longitudinal and transverse relaxation,
$\Omega$ is inhomogeneous width of the line. The system (3) is
investigated in the conditions of weakness of longitudinal fields
$(\nu_1 \ll\omega_p).$

We investigate an unperturbed system being obtained from (3) at
$\nu_1=0.$ It is easy to see that unperturbed system has two
integrals of motion:

\begin{eqnarray}
x^2+y^2+z^2=1\\
\omega_pz^2-\omega_1x=\omega_p.
\end{eqnarray}

The first of these integrals (4) corresponds to conservation of
the value of full magnetic moment, the second (5) corresponds to
the conservation of full magnetic energy. Here we assume that the
conditions of resonance are fulfilled on the unshifted frequency
$\Delta =0,$ and $x(0)=y(0)=0,~z(0)=1$ are taken as initial
conditions. These relations give definite surfaces: (4) single
sphere and (5) parabolic cylinder with generatrix along the axis
$y$. Hodographs of vector ${\bf m}$ are closed curves obtained by
crossing of these surfaces.

When $\omega_1\neq\omega_p$, hodographs are divided into two
types: 1) $\omega_1<\omega_p$ - the section consists of two
contours, which are symmetrically located on different sides of
the plane $(x,y)$; 2) $\omega_1>\omega_p$ - the section consists
of one contour twice crossing the plane $(x,y)$. When
$\omega_1=\omega_p$ the trajectory looks like symmetrical spatial
eight with self-crossing on the axis $x$, at $x=-1$ (Fig.1).
One-parametric (parameter $\omega_1$) family of eights creates the
separatrix of unperturbed motion, which divides the space into
domains according to the type of their trajectories.

\begin{figure}[!h]
\centering
\includegraphics[width=6cm]{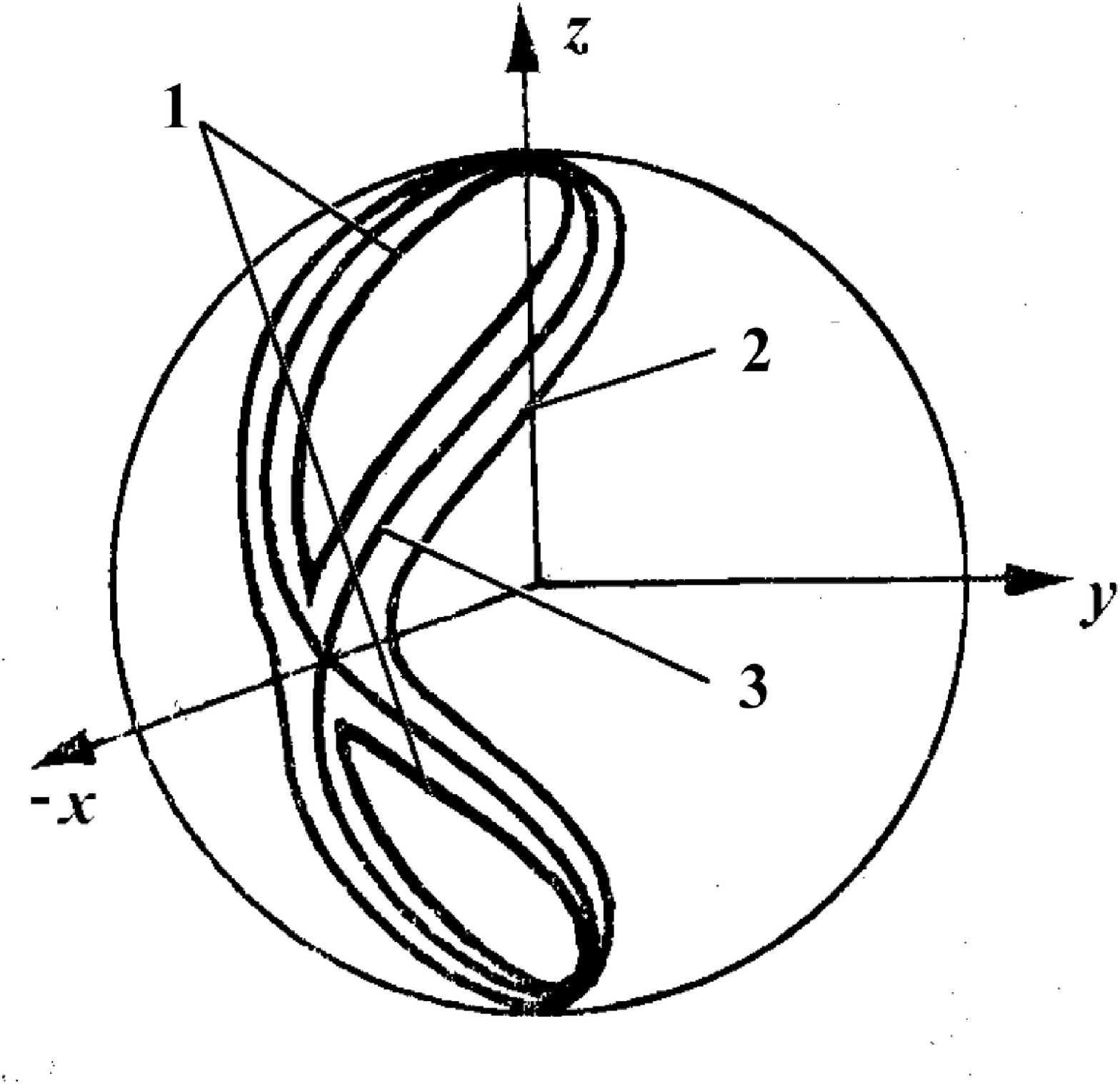}
\caption{Family of hodographs of vector ${\bf r}={\bf m}/|m_0|$ in
the conditions of resonance on the unshifted frequency: 1)
$\omega_1<\omega_p$. Two closed curves, located symmetrically
relatively to the plane $(x,y)$. 2) $\omega_1>\omega_p$. The curve
looks like the spatial eight, which crosses the plane $(x,y)$
twice. 3) $\omega_1=\omega_p$. The spatial eight with
self-crossing on the axis$x$ at $x=-1$ acts as
separatrix.}\label{fig:1}
\end{figure}

Using the integrals of motion (4) and (5) unperturbed $(\nu_1=0)$
system of equation (3) can be reduced to

\begin{equation}
\dot{z}=-\frac{1}{2}\omega_p\Biggl[-(1-z^2)\Biggl(1-\frac{\omega_1^2}{\omega_p^2}-z^2\Biggr)\Biggr]^{1/2}
\end{equation}

Solution of equation (6) can be written as:
\begin{eqnarray}
z(t^{\prime})=\left\{
\begin{array}{ll}
dn[t^{\prime};\omega_1/\omega_p]~~\omega_1<\omega_p \\
cn[t^{\prime};\omega_1/\omega_p]~~\omega_1>\omega_p
\end{array}
\right.,
\end{eqnarray}

$$t^{\prime}=-\frac{1}{2}\omega_pt,$$
where $cn$ and $dn$ are elliptical functions of Jacob, cosine and
delta of amplitude.

For the periods of motion at closed trajectories in the case of
$\omega_1<\omega_p$, we can obtain relatively
\begin{equation}
T_-=\Biggl(\frac{4}{\omega_p}\Biggr)K\Biggl(\frac{\omega_1}{\omega_p}\Biggr),
~~\omega_1<\omega_p
\end{equation}
and
\begin{equation}
T_+=\Biggl(\frac{8}{\omega_p}\Biggr)K\Biggl(\frac{\omega_p}{\omega_1}\Biggr),
~~\omega_1>\omega_p,
\end{equation}
where $K(k)$ is full elliptical integral of the first order. Hence
for the period of motion close to the separatrix
$\omega_1\approx\omega_p$ we obtain
\begin{equation}
T_c=T_+\approx2T\approx\frac{8}{\omega_1}ln\frac{4\omega_1}{|\omega_1-\omega_p|}.
\end{equation}

With the approximation to the separatrix the period of motion
logarithmically diverges. The right part of equation (3) at the
point (-1,0,0), near the knot of eight, becomes zero. It means,
the motion moderates near this point. It is evident that such
motion takes place close to separatrix hodograph (Fig.1). Thus,
close to the separatrix the motion is uneven. Near the point
(-1,0,0) it moderates, and the remaining part of hodograph is
quite rapidly overcome, so that full time of one rotation is
determined by the expression (10). Time dependence of z component
of magnetization will have the form of instanton [20]. The
$\omega_1=\omega_p$ determines the bifurcation value of amplitude
of the variable field.

Note, that at $\Delta =0$ condition of resonance is fulfilled not
at the start of motion (point (0,0,1)), but at reaching special
point (-1,0,0). It is clear that the peculiarities of the motion
near the special point are caused by fulfilling the resonance
condition in this point.

Now we discuss the case of resonance on the shifted frequency
$\Delta -\omega_p=0$. In this case the conditions of resonance are
fulfilled at the start of motion in the point (0,0,1). In NMR
experiments in the materials with big dynamic shift, resonance
condition is selected, namely, in such a way. In this case the
integrals of motion have the form:

\begin{equation}
x^2+y^2+z^2=1
\end{equation}
\begin{equation}
\omega_p(1-z)^2=\omega_1x
\end{equation}

It is clear that the crossing contours of two surfaces (11) and
(12) are topologically equitype closed curves (Fig.2), that is why
from dynamical stochasticity point of view it is not interesting.
Below we consider only the case of resonance on the unshifted
frequency $(\Delta =0)$.

\begin{figure}[!h]
\centering
\includegraphics[width=6cm]{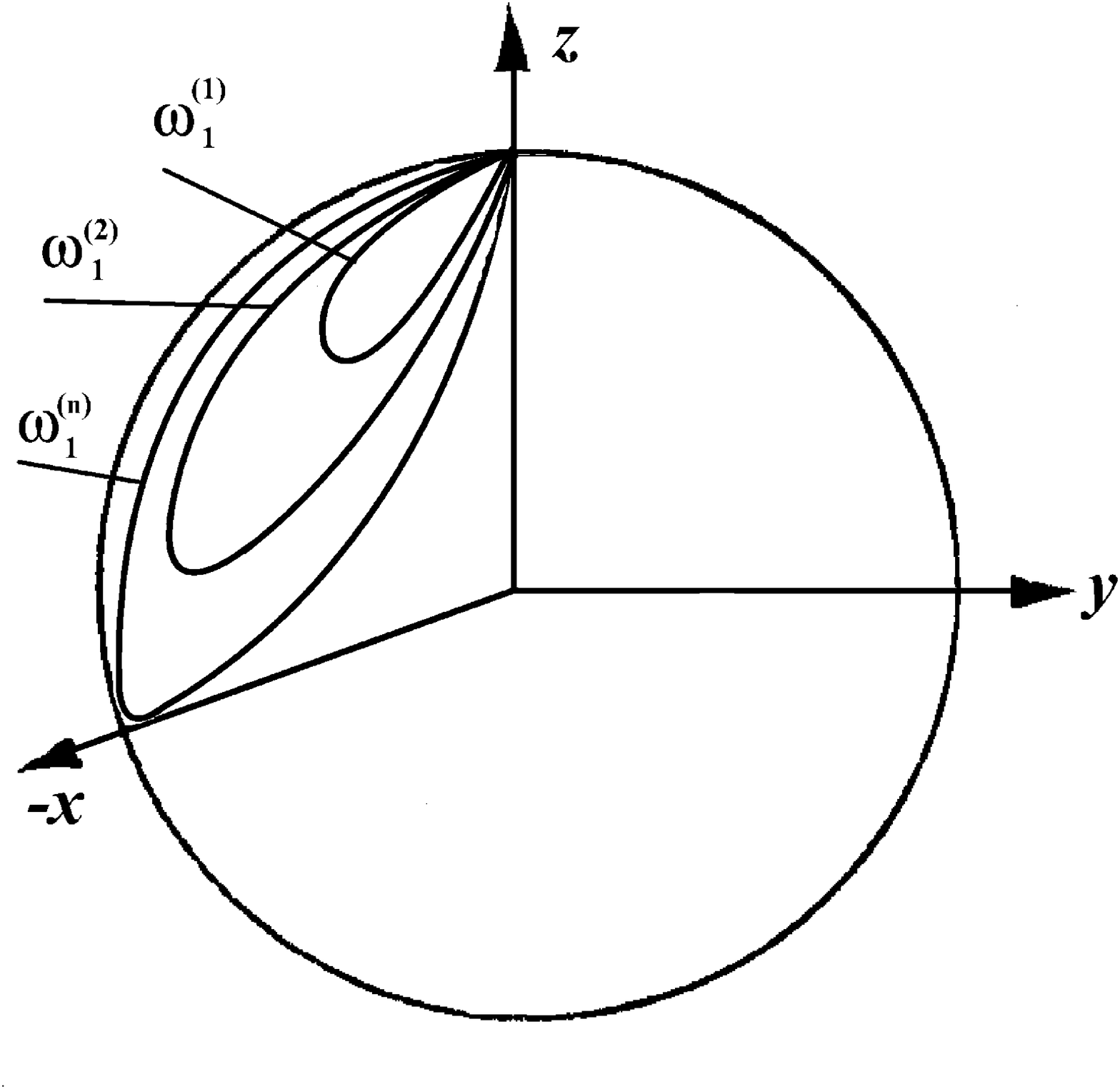}
\caption{The family of hodographs of vector ${\bf r}={\bf m}/|m_0|$ in the
conditions of resonance on the shifted frequency $\Delta
-\omega_p=0$. With the growth of
$\omega_1(\omega_1^{(1)}<\omega_1^{(2)}<\ldots \omega_1^{(n)})$
the hodographs increase in dimensions and in the limit
$\omega_1\approx\omega_p$ reach the point (-1,00).}\label{fig:2}
\end{figure}

\section{\textbf{Analysis of Results of Computer Modeling}}
\label{3} For analysis of the above mentioned discussions and
investigation of the influence of perturbation on motion of
nuclear magnetization the initial equations (1) describing
spin-system dynamics were modeled. Modeling was made in the medium
of Simulink of mathematical package MATLAB. During the modeling we
used stroboscopic method, according to which observation of vector
motion of magnetization was conducted not continuously, but during
separate, short periodically following one another time intervals.

\begin{figure}[!h]
\centering \includegraphics[width=11cm]{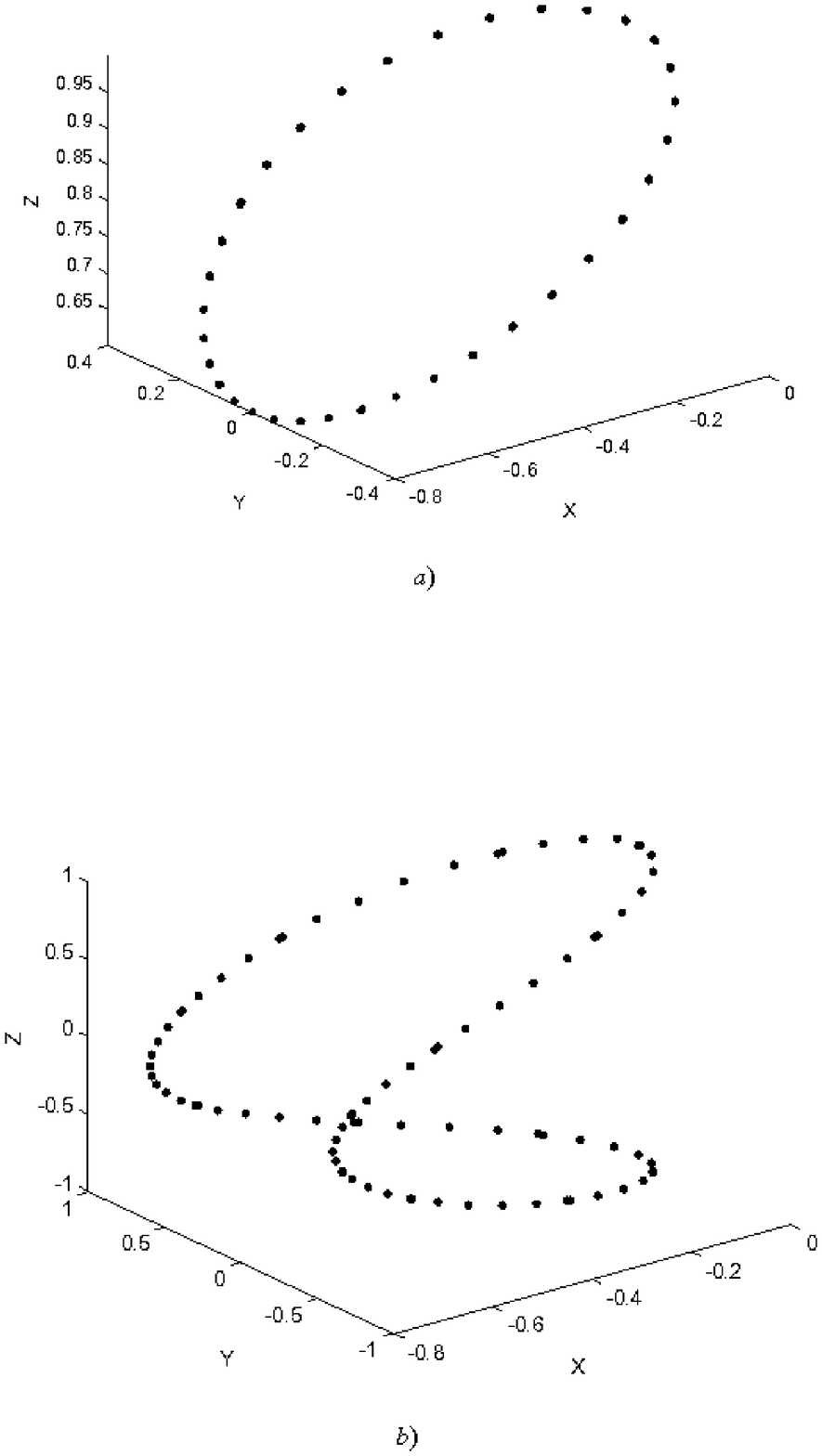}
\caption{Stroboscopic points of hodographs, which are far from
separatrix $(\omega_1\neq\omega_p).~a)~ \omega_1=800Hz,~
\omega_p=1000Hz,~\nu_1=0;~~b)~\omega_1=1000Hz,~
\omega_p=800Hz,~\nu_1=0$. In both cases the set of stroboscopic
points is located periodically.}\label{fig:3}
\end{figure}

\begin{figure}[!h]
\centering \includegraphics[width=11cm]{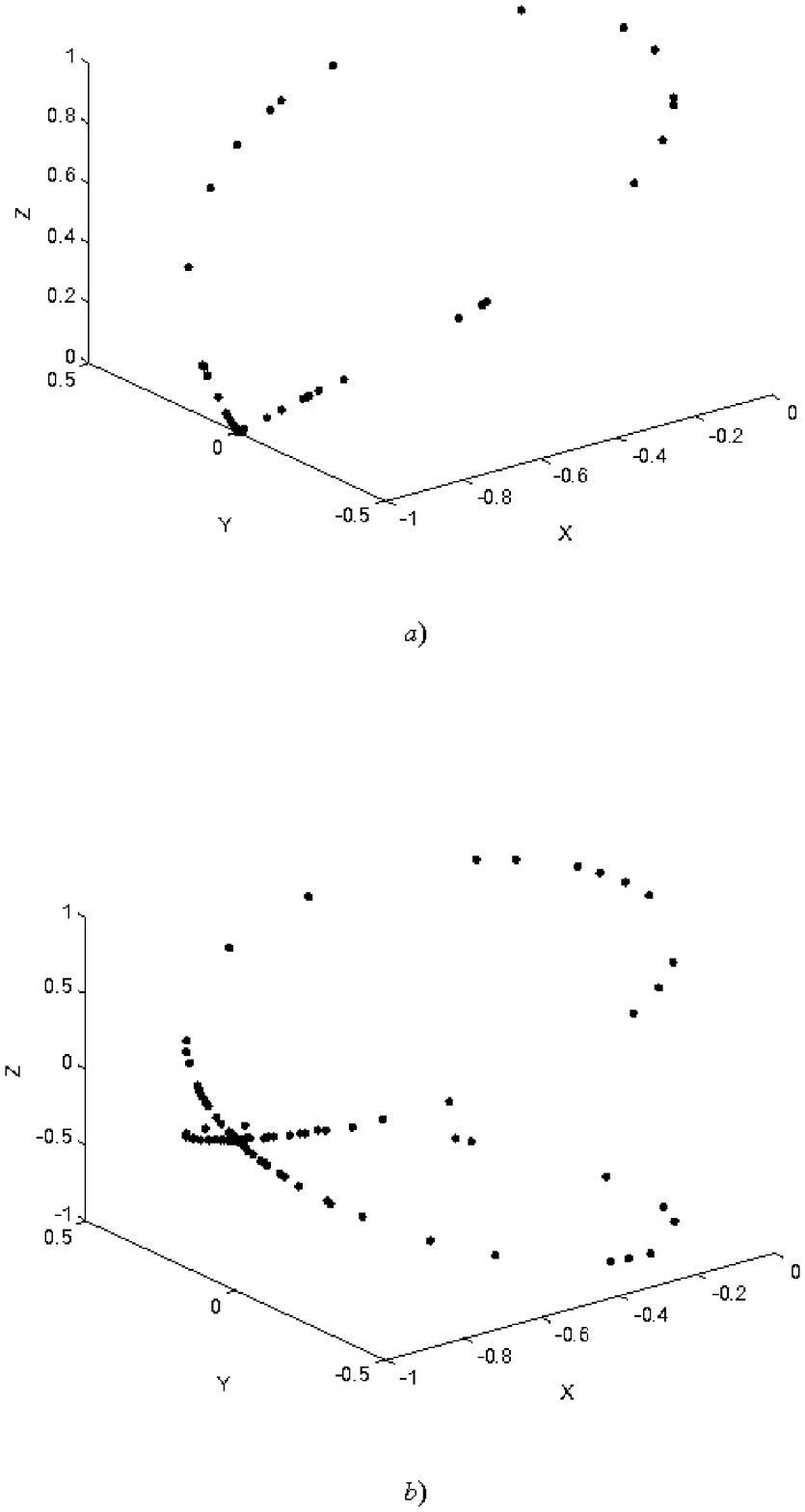}
\caption{Stroboscopic points of separatrix hodograph: a)
$\omega_1=\omega_p=1000Hz,~\nu_1=0$. Chaotic points located on the
upper part of hodograph. b)
$\omega_1=\omega_p=1000Hz,~\nu_1=10^{-4}Hz$. Chaotic points
located along the whole separatrix hodograph.}\label{fig:4}
\end{figure}

Fig.3 shows hodographs distant from separatrix
$(\omega_1\neq\omega_p)$. The points being set using stroboscopic
method are periodically located along the hodograph. Approaching
the separatrix $(\omega_1\rightarrow\omega_p)$ they become chaotic
(Fig.4). In the case of $\omega_1\approx\omega_p$ and $\nu_1=0$
chaotic points cover only the upper part of separatrix eight, but
after switching on perturbation ($\nu_1=10^{-4}Hz$) the motion
acts along the whole separatrix one. Stroboscopic points of
hodograph are chaotic in this case too. Consequently, a role of
slowly changing perturbation is reduced to regular variation of
the place of the point of self-crossing separatrix eight.

Thus, stochastic set of points origins close to the singular point
(-1,0,0), overcoming of which the motion becomes unpredictable:
all four directions, crossing the special point become equally
probable. Notice that set of points turned out to be stochastic
though external influence on the system is periodical.

As a result of numerical integration of (3) for  $z(t)$ at the
motion close to the separatrix the solution in the form of
periodical succession of instantons is obtained (Fig.5). In the
case of $\omega_1-\omega_p\rightarrow-0$ all the instantons are
positive, while at $\omega_1-\omega_p\rightarrow+0$ in the
periodical succession alternate change of instanton sign takes
place.

The same result can be obtained also by means of direct set of
figures of elliptical functions of Jacob (7).

\begin{figure}[!h]
\centering
\includegraphics[width=15cm]{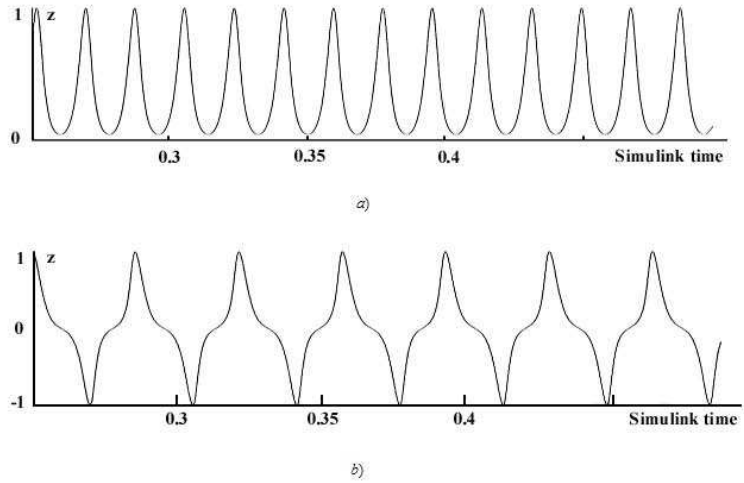} \caption{Change
of $z$ component of nuclear magnetization during simulation for
separatrix close to trajectory
$a)~\omega_p\geq\omega_1~(\omega_1=1000Hz,~\omega_p=1001Hz,~\nu_1=0);~~b)~\omega_p\leq\omega_1~
(\omega_1=1001Hz,~\omega_p=1000Hz,~\nu_1=0)$.}\label{fig:5}
\end{figure}

\begin{figure}[!h]
\centering
\includegraphics[width=15cm]{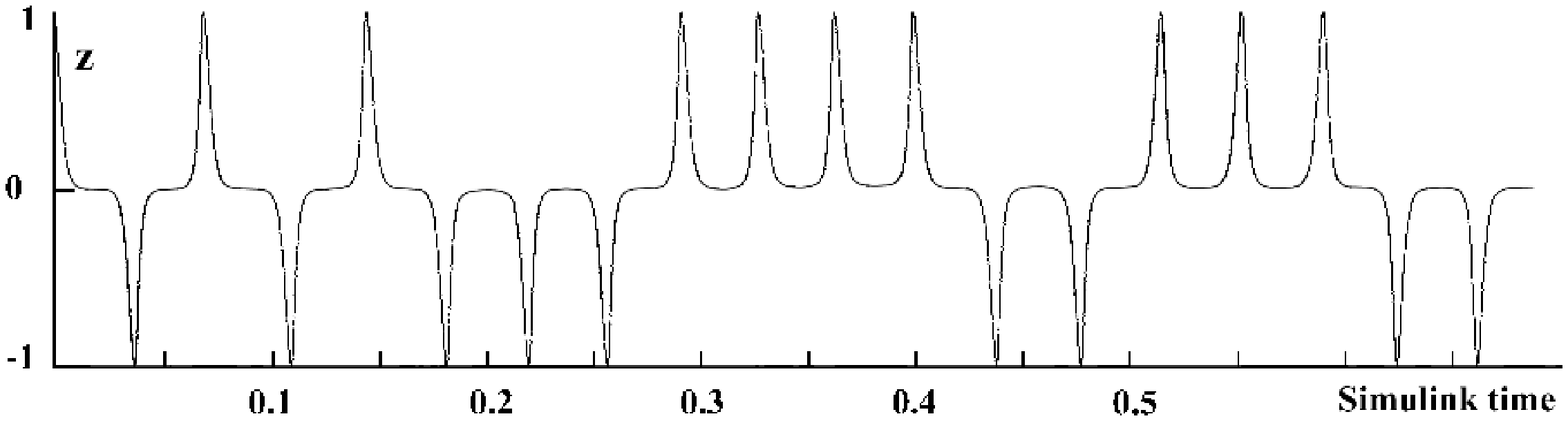} \caption{Change
of $z$ component of vector of nuclear magnetization during
simulation for separatrix trajectory
$(\omega_p=\omega_1=1000Hz,~\nu_1=10^{-4}Hz)$.}\label{fig:6}
\end{figure}

Fig.6 shows the fragment of the picture of changing the sign of
$m_z$ during simulation already at the switched periodical
perturbation ($\nu_1=10^{-4}Hz$). Disordered change of the sign of
nuclear magnetization testifies an origin of stochasticity at the
junction over branching point.

Thus, hodograph of vector of nuclear magnetization at nonlinear
NMR on the unshifted frequency in the conditions
$\omega_1\approx\omega_p$ possess the properties of separatrix.
This is the basis for appearance of the set of chaotic points in
the stroboscopic picture of separatrix hodograph (Fig.3), and also
for chaotic change of instantons signs of $z$ component of
magnetization (Fig.6).\\

\textbf{Acknowledgement}\\

The designated project has been fulfilled by financial support
from the Georgian National Foundation (grants: GNSF/STO 7/4-197,
GNSF/STO 7/4-179). The financial support of Deutsche
Forschungsgemeinschaft through SPP 1285 (contract number EC94/5-1)
is gratefully acknowledged by L. Chotorlishvili.



\end{document}